\documentclass[twocolumn,letterpaper,aps,prl,longbibliography,superscriptaddress,floatfix]{revtex4-1}

\usepackage{graphicx}	
\usepackage{xspace}	

\newcommand{\s}[1]{\mbox{$\sqrt{s}$ = #1\,GeV}}
\newcommand{\snn}[1]{\mbox{$\sqrt{s_{_{NN}}}$ = #1\,GeV}}
 
\newcommand{\pp}{$p$$+$$p$ }
\newcommand{\AB}[2]{#1$+$#2}

\newcommand{\pt}{\mbox{$p_T$}\xspace}

\newcommand{\ee}{\mbox{$e^+e^-$}}

\newcommand{\pte}{\mbox{$p_T^{ee}$}\xspace}

\newcommand{\Npart}{\mbox{$N_{\rm part}$}\xspace}
\newcommand{\Ncoll}{\mbox{$N_{\rm coll}$}\xspace}
\newcommand{\Nch}{\mbox{$N_{\rm ch}$}\xspace}
\newcommand{\dNch}{\mbox{$dN_{\rm ch}/d\eta$}\xspace}

\newcommand{\sqs}{\mbox{$\sqrt{s}$}\xspace}
\newcommand{\sqsn}{\mbox{$\sqrt{s_{_{NN}}}$}\xspace}

\newcommand{\Rg}{\mbox{$R_\gamma$}\xspace}
\newcommand{\Nincl}{\mbox{$N_\gamma^{\rm incl}$}\xspace}
\newcommand{\Ntag} {\mbox{$N_\gamma^{\pi^0,{\rm tag}}$}\xspace}
\newcommand{\ef}{\mbox{$\langle\varepsilon_{\gamma} f \rangle$}\xspace}
\newcommand{\gevc}{\mbox{GeV/$c$}\xspace}
\newcommand{\mevc}{\mbox{MeV/$c$}\xspace}

\begin{document}

\title{Beam-energy and centrality dependence of direct-photon emission 
from ultra-relativistic heavy-ion collisions}

\newcommand{\abilene}{Abilene Christian University, Abilene, Texas 79699, USA}
\newcommand{\acadsin}{Institute of Physics, Academia Sinica, Taipei 11529, Taiwan}
\newcommand{\augie}{Department of Physics, Augustana University, Sioux Falls, South Dakota 57197, USA}
\newcommand{\banaras}{Department of Physics, Banaras Hindu University, Varanasi 221005, India}
\newcommand{\barc}{Bhabha Atomic Research Centre, Bombay 400 085, India}
\newcommand{\baruch}{Baruch College, City University of New York, New York, New York, 10010 USA}
\newcommand{\bnlcoll}{Collider-Accelerator Department, Brookhaven National Laboratory, Upton, New York 11973-5000, USA}
\newcommand{\bnlphys}{Physics Department, Brookhaven National Laboratory, Upton, New York 11973-5000, USA}
\newcommand{\caucr}{University of California-Riverside, Riverside, California 92521, USA}
\newcommand{\charlesczech}{Charles University, Ovocn\'{y} trh 5, Praha 1, 116 36, Prague, Czech Republic}
\newcommand{\chonbuk}{Chonbuk National University, Jeonju, 561-756, Korea}
\newcommand{\ciae}{Science and Technology on Nuclear Data Laboratory, China Institute of Atomic Energy, Beijing 102413, People's Republic of China}
\newcommand{\cns}{Center for Nuclear Study, Graduate School of Science, University of Tokyo, 7-3-1 Hongo, Bunkyo, Tokyo 113-0033, Japan}
\newcommand{\colorado}{University of Colorado, Boulder, Colorado 80309, USA}
\newcommand{\columbia}{Columbia University, New York, New York 10027 and Nevis Laboratories, Irvington, New York 10533, USA}
\newcommand{\czechtech}{Czech Technical University, Zikova 4, 166 36 Prague 6, Czech Republic}
\newcommand{\dapnia}{Dapnia, CEA Saclay, F-91191, Gif-sur-Yvette, France}
\newcommand{\debrecen}{Debrecen University, H-4010 Debrecen, Egyetem t{\'e}r 1, Hungary}
\newcommand{\elte}{ELTE, E{\"o}tv{\"o}s Lor{\'a}nd University, H-1117 Budapest, P{\'a}zm{\'a}ny P.~s.~1/A, Hungary}
\newcommand{\eszterhazy}{Eszterh\'azy K\'aroly University, K\'aroly R\'obert Campus, H-3200 Gy\"ongy\"os, M\'atrai \'ut 36, Hungary}
\newcommand{\ewha}{Ewha Womans University, Seoul 120-750, Korea}
\newcommand{\fit}{Florida Institute of Technology, Melbourne, Florida 32901, USA}
\newcommand{\fsu}{Florida State University, Tallahassee, Florida 32306, USA}
\newcommand{\gsu}{Georgia State University, Atlanta, Georgia 30303, USA}
\newcommand{\hanyang}{Hanyang University, Seoul 133-792, Korea}
\newcommand{\hiroshima}{Hiroshima University, Kagamiyama, Higashi-Hiroshima 739-8526, Japan}
\newcommand{\howard}{Department of Physics and Astronomy, Howard University, Washington, DC 20059, USA}
\newcommand{\ihepprot}{IHEP Protvino, State Research Center of Russian Federation, Institute for High Energy Physics, Protvino, 142281, Russia}
\newcommand{\illuiuc}{University of Illinois at Urbana-Champaign, Urbana, Illinois 61801, USA}
\newcommand{\inrras}{Institute for Nuclear Research of the Russian Academy of Sciences, prospekt 60-letiya Oktyabrya 7a, Moscow 117312, Russia}
\newcommand{\instpasczech}{Institute of Physics, Academy of Sciences of the Czech Republic, Na Slovance 2, 182 21 Prague 8, Czech Republic}
\newcommand{\isu}{Iowa State University, Ames, Iowa 50011, USA}
\newcommand{\jaea}{Advanced Science Research Center, Japan Atomic Energy Agency, 2-4 Shirakata Shirane, Tokai-mura, Naka-gun, Ibaraki-ken 319-1195, Japan}
\newcommand{\jinrdubna}{Joint Institute for Nuclear Research, 141980 Dubna, Moscow Region, Russia}
\newcommand{\jyvaskyla}{Helsinki Institute of Physics and University of Jyv{\"a}skyl{\"a}, P.O.Box 35, FI-40014 Jyv{\"a}skyl{\"a}, Finland}
\newcommand{\kaeri}{KAERI, Cyclotron Application Laboratory, Seoul, Korea}
\newcommand{\kek}{KEK, High Energy Accelerator Research Organization, Tsukuba, Ibaraki 305-0801, Japan}
\newcommand{\korea}{Korea University, Seoul, 02841, Korea}
\newcommand{\kurchatov}{National Research Center ``Kurchatov Institute", Moscow, 123098 Russia}
\newcommand{\kyoto}{Kyoto University, Kyoto 606-8502, Japan}
\newcommand{\labllr}{Laboratoire Leprince-Ringuet, Ecole Polytechnique, CNRS-IN2P3, Route de Saclay, F-91128, Palaiseau, France}
\newcommand{\lahorelums}{Physics Department, Lahore University of Management Sciences, Lahore 54792, Pakistan}
\newcommand{\lawllnl}{Lawrence Livermore National Laboratory, Livermore, California 94550, USA}
\newcommand{\losalamos}{Los Alamos National Laboratory, Los Alamos, New Mexico 87545, USA}
\newcommand{\lpc}{LPC, Universit{\'e} Blaise Pascal, CNRS-IN2P3, Clermont-Fd, 63177 Aubiere Cedex, France}
\newcommand{\lund}{Department of Physics, Lund University, Box 118, SE-221 00 Lund, Sweden}
\newcommand{\lyon}{IPNL, CNRS/IN2P3, Univ Lyon, Université Lyon 1, F-69622, Villeurbanne, France}
\newcommand{\maryland}{University of Maryland, College Park, Maryland 20742, USA}
\newcommand{\mass}{Department of Physics, University of Massachusetts, Amherst, Massachusetts 01003-9337, USA}
\newcommand{\michigan}{Department of Physics, University of Michigan, Ann Arbor, Michigan 48109-1040, USA}
\newcommand{\muenster}{Institut f\"ur Kernphysik, University of M\"unster, D-48149 M\"unster, Germany}
\newcommand{\muhlenberg}{Muhlenberg College, Allentown, Pennsylvania 18104-5586, USA}
\newcommand{\myongji}{Myongji University, Yongin, Kyonggido 449-728, Korea}
\newcommand{\nagasaki}{Nagasaki Institute of Applied Science, Nagasaki-shi, Nagasaki 851-0193, Japan}
\newcommand{\nara}{Nara Women's University, Kita-uoya Nishi-machi Nara 630-8506, Japan}
\newcommand{\natmephi}{National Research Nuclear University, MEPhI, Moscow Engineering Physics Institute, Moscow, 115409, Russia}
\newcommand{\newmex}{University of New Mexico, Albuquerque, New Mexico 87131, USA}
\newcommand{\nmsu}{New Mexico State University, Las Cruces, New Mexico 88003, USA}
\newcommand{\ohio}{Department of Physics and Astronomy, Ohio University, Athens, Ohio 45701, USA}
\newcommand{\ornl}{Oak Ridge National Laboratory, Oak Ridge, Tennessee 37831, USA}
\newcommand{\orsay}{IPN-Orsay, Univ.~Paris-Sud, CNRS/IN2P3, Universit\'e Paris-Saclay, BP1, F-91406, Orsay, France}
\newcommand{\peking}{Peking University, Beijing 100871, People's Republic of China}
\newcommand{\pnpi}{PNPI, Petersburg Nuclear Physics Institute, Gatchina, Leningrad region, 188300, Russia}
\newcommand{\riken}{RIKEN Nishina Center for Accelerator-Based Science, Wako, Saitama 351-0198, Japan}
\newcommand{\rikjrbrc}{RIKEN BNL Research Center, Brookhaven National Laboratory, Upton, New York 11973-5000, USA}
\newcommand{\rikkyo}{Physics Department, Rikkyo University, 3-34-1 Nishi-Ikebukuro, Toshima, Tokyo 171-8501, Japan}
\newcommand{\saispbstu}{Saint Petersburg State Polytechnic University, St.~Petersburg, 195251 Russia}
\newcommand{\saopaulo}{Universidade de S{\~a}o Paulo, Instituto de F\'{\i}sica, Caixa Postal 66318, S{\~a}o Paulo CEP05315-970, Brazil}
\newcommand{\seoulnat}{Department of Physics and Astronomy, Seoul National University, Seoul 151-742, Korea}
\newcommand{\stonybrkc}{Chemistry Department, Stony Brook University, SUNY, Stony Brook, New York 11794-3400, USA}
\newcommand{\stonycrkp}{Department of Physics and Astronomy, Stony Brook University, SUNY, Stony Brook, New York 11794-3800, USA}
\newcommand{\subatech}{SUBATECH (Ecole des Mines de Nantes, CNRS-IN2P3, Universit{\'e} de Nantes) BP 20722-44307, Nantes, France}
\newcommand{\sungskku}{Sungkyunkwan University, Suwon, 440-746, Korea}
\newcommand{\tenn}{University of Tennessee, Knoxville, Tennessee 37996, USA}
\newcommand{\titech}{Department of Physics, Tokyo Institute of Technology, Oh-okayama, Meguro, Tokyo 152-8551, Japan}
\newcommand{\tsukuba}{Tomonaga Center for the History of the Universe, University of Tsukuba, Tsukuba, Ibaraki 305, Japan}
\newcommand{\vandy}{Vanderbilt University, Nashville, Tennessee 37235, USA}
\newcommand{\waseda}{Waseda University, Advanced Research Institute for Science and Engineering, 17  Kikui-cho, Shinjuku-ku, Tokyo 162-0044, Japan}
\newcommand{\weizmann}{Weizmann Institute, Rehovot 76100, Israel}
\newcommand{\wigner}{Institute for Particle and Nuclear Physics, Wigner Research Centre for Physics, Hungarian Academy of Sciences (Wigner RCP, RMKI) H-1525 Budapest 114, POBox 49, Budapest, Hungary}
\newcommand{\yonsei}{Yonsei University, IPAP, Seoul 120-749, Korea}
\newcommand{\zagreb}{Department of Physics, Faculty of Science, University of Zagreb, Bijeni\v{c}ka c.~32 HR-10002 Zagreb, Croatia}
\affiliation{\abilene}
\affiliation{\acadsin}
\affiliation{\augie}
\affiliation{\banaras}
\affiliation{\barc}
\affiliation{\baruch}
\affiliation{\bnlcoll}
\affiliation{\bnlphys}
\affiliation{\caucr}
\affiliation{\charlesczech}
\affiliation{\chonbuk}
\affiliation{\ciae}
\affiliation{\cns}
\affiliation{\colorado}
\affiliation{\columbia}
\affiliation{\czechtech}
\affiliation{\dapnia}
\affiliation{\debrecen}
\affiliation{\elte}
\affiliation{\eszterhazy}
\affiliation{\ewha}
\affiliation{\fit}
\affiliation{\fsu}
\affiliation{\gsu}
\affiliation{\hanyang}
\affiliation{\hiroshima}
\affiliation{\howard}
\affiliation{\ihepprot}
\affiliation{\illuiuc}
\affiliation{\inrras}
\affiliation{\instpasczech}
\affiliation{\isu}
\affiliation{\jaea}
\affiliation{\jinrdubna}
\affiliation{\jyvaskyla}
\affiliation{\kaeri}
\affiliation{\kek}
\affiliation{\korea}
\affiliation{\kurchatov}
\affiliation{\kyoto}
\affiliation{\labllr}
\affiliation{\lahorelums}
\affiliation{\lawllnl}
\affiliation{\losalamos}
\affiliation{\lpc}
\affiliation{\lund}
\affiliation{\lyon}
\affiliation{\maryland}
\affiliation{\mass}
\affiliation{\michigan}
\affiliation{\muenster}
\affiliation{\muhlenberg}
\affiliation{\myongji}
\affiliation{\nagasaki}
\affiliation{\nara}
\affiliation{\natmephi}
\affiliation{\newmex}
\affiliation{\nmsu}
\affiliation{\ohio}
\affiliation{\ornl}
\affiliation{\orsay}
\affiliation{\peking}
\affiliation{\pnpi}
\affiliation{\riken}
\affiliation{\rikjrbrc}
\affiliation{\rikkyo}
\affiliation{\saispbstu}
\affiliation{\saopaulo}
\affiliation{\seoulnat}
\affiliation{\stonybrkc}
\affiliation{\stonycrkp}
\affiliation{\subatech}
\affiliation{\sungskku}
\affiliation{\tenn}
\affiliation{\titech}
\affiliation{\tsukuba}
\affiliation{\vandy}
\affiliation{\waseda}
\affiliation{\weizmann}
\affiliation{\wigner}
\affiliation{\yonsei}
\affiliation{\zagreb}
\author{A.~Adare} \affiliation{\colorado} 
\author{S.~Afanasiev} \affiliation{\jinrdubna} 
\author{C.~Aidala} \affiliation{\columbia} \affiliation{\losalamos} \affiliation{\mass} \affiliation{\michigan} 
\author{N.N.~Ajitanand} \altaffiliation{Deceased} \affiliation{\stonybrkc} 
\author{Y.~Akiba} \email[PHENIX Spokesperson: ]{akiba@rcf.rhic.bnl.gov} \affiliation{\riken} \affiliation{\rikjrbrc} 
\author{R.~Akimoto} \affiliation{\cns} 
\author{H.~Al-Bataineh} \affiliation{\nmsu} 
\author{J.~Alexander} \affiliation{\stonybrkc} 
\author{M.~Alfred} \affiliation{\howard} 
\author{A.~Al-Jamel} \affiliation{\nmsu} 
% ...
\collaboration{PHENIX Collaboration} \noaffiliation

\date{\today}


\begin{abstract}

The PHENIX collaboration presents first measurements of low-momentum 
($0.4<p_T<3$ GeV/$c$) direct-photon yields from Au$+$Au collisions at 
$\sqrt{s_{_{NN}}}$=39 and 62.4 GeV.  For both beam energies the 
direct-photon yields are substantially enhanced with respect to 
expectations from prompt processes, similar to the yields observed in 
Au$+$Au collisions at $\sqrt{s_{_{NN}}}$=200.  Analyzing the photon 
yield as a function of the experimental observable $dN_{\rm ch}/d\eta$ 
reveals that the low-momentum ($>$1\,GeV/$c$) direct-photon yield 
$dN_{\gamma}^{\rm dir}/d\eta$ is a smooth function of 
$dN_{\rm ch}/d\eta$ and can be well described as proportional to 
$(dN_{\rm ch}/d\eta)^\alpha$ with $\alpha{\approx}1.25$. This scaling 
behavior holds for a wide range of beam energies at the Relativistic
Heavy Ion Collider and the Large Hadron Collider, for
centrality selected samples, as well as for different, $A$$+$$A$ 
collision systems. At a given beam energy the scaling also holds for 
high $p_T$ ($>5$\,GeV/$c$) but when results from different collision 
energies are compared, an additional $\sqrt{s_{_{NN}}}$-dependent 
multiplicative factor is needed to describe the integrated-direct-photon 
yield.  

\end{abstract}

\maketitle


Measurements of direct photons provide information about the 
strongly coupled quark-gluon plasma (QGP) produced in heavy ion 
collisions and its ``fireball" evolution to hadron resonance 
matter. Due to their long mean free path photons do not interact 
with the matter and thus their spectra provide information about 
all stages of the collision integrated over space and 
time~\cite{Stankus:2005eq,David:2006sr,Linnyk:2015rco}. In 
particular low \pt photons in the momentum range up to a few 
GeV/$c$ are expected to carry information about the hot and dense 
fireball.

In experiments direct photons are detected simultaneously with a 
much larger number of photons from hadron decays, mostly from 
$\pi^0$ and $\eta$ mesons. The main challenge is to subtract these 
decay contributions from the measurement to obtain the photons 
directly emitted from the collision. In addition to photons from 
the hot fireball, direct photons include those emitted from initial 
hard scattering processes, such as quark-gluon Compton scattering 
among the incoming partons~\cite{Fritzsch:1977eq}. Disentangling 
this prompt component from the photons emitted from the fireball is 
an additional challenge.

First evidence for direct photon emission from heavy ion collisions 
came from WA98 \cite{Aggarwal:2003zy,Aggarwal:2000th}, with 
conclusive results only for $p_T > 1.5$ \gevc. PHENIX established 
that a large number of low \pt direct photons are radiated from the 
fireball created in \AB{Au}{Au} collisions at \snn{200} 
\cite{Adare:2008ab} and that their yield increases with a power of 
\Npart while the inverse slopes of the spectra are independent of 
the centrality of the collisions \cite{Adare:2014fwh}. 
Simultaneously, low \pt direct photon emission exhibits a 
significant azimuthal anisotropy with respect to the reaction 
plane~\cite{Adare:2011zr,Adare:2015lcd}.

ALICE has published \cite{Adam:2015lda,Lohner:2012ct} similar 
observations of low \pt direct photons from \AB{Pb}{Pb} collisions 
at \snn{2760}. STAR also reported a measurement of the direct 
photon yields in \AB{Au}{Au} at \snn{200} \cite{STAR:2016use}, the 
published yields are significantly lower compared to PHENIX 
results. The origin of the discrepancy remains 
unresolved~\footnote{We note that PHENIX has published consistent 
results from several independent analyses with different methods, 
using virtual photons \protect\cite{Adare:2008ab}, which is the 
method adopted by STAR \protect\cite{STAR:2016use}, and using 
photon conversions in the detector material 
\protect\cite{Adare:2014fwh}. A third method using photons measured 
through their energy deposited in the electromagnetic calorimeter 
to reconstruct low \pt photons has not been published 
\protect\cite{GongPhD}, but gives consistent results as well.}.

A large body of theoretical work on low \pt direct photon emission 
in \AB{A}{A} collisions exists in the literature. Many model 
calculations are qualitatively consistent with the data, but a 
quantitative description remains difficult, primarily due to the 
simultaneous observation of large yields and large azimuthal 
anisotropies \cite{vanHees:2011vb, vanHees:2014ida, Dion:2011pp, 
Shen:2013vja, Shen:2015qba, Paquet:2015lta, Bratkovskaya:2008iq, 
Bratkovskaya:2014mva, Linnyk:2013wma, Chiu:2012ij, 
McLerran:2014hza, McLerran:2015mda, Berges:2017eom, monnai:2014kqa, 
Lee:2014pwa, Turbide:2003si, Dusling:2009ej, Lee:2014pwa, 
Heffernan:2014mla, Linnyk:2015tha, Basar:2012bp, Basar:2014swa, 
Muller:2013ila, Ayala:2017vex,Goloviznin:2018mwb}.

To provide further insights, PHENIX is investigating the system 
size dependence of direct photon emission from heavy ion collisions 
by varying beam energy, centrality, and collision species. In this 
publication we present low-\pt direct photon data from \AB{Au}{Au} 
collisions at \snn{39} and 62.4 GeV taken with the PHENIX 
experiment in 2010.  We compare the centrality selected spectra and 
integrated yields from \AB{Au}{Au} to those from \pp collisions at 
\snn{200} \cite{Adare:2008ab, Adare:2014fwh}, \AB{Cu}{Cu} 
collisions at \snn{200} \cite{Adare:2018jsz}, and \AB{Pb}{Pb} 
collisions at \snn{2760} \cite{Adam:2015lda}. This study covers a 
factor of 70 in \sqsn and nearly two orders of magnitude in system 
size.

The 39 and 62.4 GeV direct photon spectra are obtained from two 
data samples of minimum-bias (MB) \AB{Au}{Au} collisions that have 
a total of $7.79{\times}10^7$ and $2.12{\times}10^8$ events, 
respectively. The MB trigger and centrality selection is derived 
from data taken with the PHENIX beam-beam 
counters~\cite{Allen:2003zt}. The data analysis uses the same 
techniques deployed for the analysis of the \snn{200} \AB{Au}{Au} 
data \cite{Adare:2014fwh}, which were taken in the same year under 
nearly identical conditions. Here we give a brief overview of the 
setup and data analysis, and refer to our previous publication for 
more details \cite{Adare:2014fwh}.

Photons are reconstructed through their conversion to \ee pairs in 
the detector material, specifically the readout boards of the 
hadron blind detector (HBD) \cite{Anderson:2011jw} that are located 
at a radius of $60\,{\rm cm}$ from the beam axes. The trajectories 
and momenta of the $e^+$ and $e^-$ are determined by the central 
arm tracking detectors~\cite{Adcox:2003zp}. Each of the two central 
arms covers 90$^\circ$ in azimuth and a rapidity range of 
$|\eta|<0.35$. A transverse momentum cut, \pt $>200$ MeV/c, is 
applied to each trajectory.  To identify trajectories as $e^+$ or 
$e^-$ candidates, we require a minimum of three associated signals 
in the ring-imaging \v{C}erenkov detector~\cite{Aizawa:2003zq} and 
that the energy measured in the electromagnetic calorimeter 
(EMCal)~\cite{Aphecetche:2003zr} matches the measured momentum 
($E/p>0.5$).

All $e^+$ and $e^-$ reconstructed in the same arm are matched to 
pairs. In the 2010 setup there is no tracking near the collision 
point, so the origin of an individual track is unknown. Thus, for 
each \ee pair the mass is calculated twice: first assuming the pair 
originated at the event vertex ($m_{\rm vtx}$), then assuming the 
\ee \ is a conversion pair from the HBD readout boards $(m_{\rm 
HBD})$. In the latter case, $m_{\rm HBD}$ will be consistent with 
zero, within a mass resolution of a few MeV/c$^2$, while $m_{\rm 
vtx}$ will be about 12 MeV/c$^2$. With a cut on both masses a 
sample of photon conversion is selected with a purity of about 
99\%. The combinatorial background is negligible, because the 
conversion material, in radiation length $X/X_0{\approx}3\%$, is 
about 10 times thicker than materials closer to the vertex; and it 
is at a relatively large distance from the event vertex. The 1\% 
contamination is mostly from $\pi^0$ Dalitz decays, $\pi^{0} 
\rightarrow \gamma e^{+}e^{-}$, and from conversions in front of 
the HBD readout boards.

The direct photon content in the photon sample is determined by the 
ratio $R_\gamma$, which is the ratio of all emitted photons 
($\gamma^{\rm incl}$) to those from hadron decays ($\gamma^{\rm 
hadron}$). The ratio $R_{\gamma}$ is determined from a double 
ratio:
\begin{equation}
R_{\gamma} = \frac{\gamma^{\rm incl}}{\gamma^{\rm hadron}} = \frac{\langle\varepsilon_{\gamma} f \rangle \left(N_{\gamma}^{\rm incl} / N_{\gamma}^{\pi^{0},{\rm tag}}\right)_{\rm Data}}{\left(\gamma^{\rm hadron} /  \gamma^{\pi^{0}}\right)_{\rm Sim}}.
\label{eqn:rgamma}
\end{equation}

All quantities in this double ratio are functions of the conversion 
photon \pte. The measured quantities are the number of detected 
conversion photons \Nincl and the subset of those that are tagged 
as $\pi^0$ decay photon \Ntag. The tagged photons \Ntag are 
determined statistically in bins of the \pte. Each conversion 
photon is paired with all showers with $E >$ 400 MeV measured in 
the EMCal of the same arm. The invariant $e^{+}e^{-}\gamma$ mass is 
calculated and the counts above the combinatorial background in the 
$\pi^0$ mass peak give \Ntag. To convert the ratio \Nincl/\Ntag to 
$\gamma^{\rm incl}/\gamma^{\pi^0}$ only \Ntag needs to be corrected 
for the momentum averaged conditional acceptance-efficiency \ef 
that the second decay photon can be reconstructed in the EMCal. All 
other corrections to the numerator and denominator 
cancel~\cite{Adare:2014fwh}. Because rather loose cuts are applied 
to the EMCal showers, \ef is mostly determined by the $\pi^0$ decay 
kinematics, the detector geometry, and the energy cut. Thus, \ef 
can be calculated to a few percent accuracy using a Monte-Carlo 
simulation of $\pi^0$ decays. Photons from pions are determined 
from the measured $\pi^0$ spectra \cite{Adare:2012uk} and two body 
decay kinematics. The spectrum of decay photons ($\gamma^{\rm 
hadron}$) is derived from $\gamma^{\pi^0}$ and the $\eta/\pi^0$ 
ratio~\cite{Adare:2012wg}, which is independent of collision system 
and energy, with additional contribution from heavier mesons of 
about 4\%.

Once \Rg is established, the direct photon spectrum can be 
calculated as:
\begin{equation}
\gamma^{\rm direct} = (R_{\gamma} - 1)\ \gamma^{\rm hadron}.
\label{eqn:yield}
\end{equation}

The uncertainty on $\gamma^{\rm hadron}$, approximately 
10\%~\cite{Adare:2014fwh}, cancels in \Rg (with that of 
$\gamma^{\pi^{0}}$ in Eq.~(\ref{eqn:rgamma})) but has to be applied 
to $\gamma^{\rm direct}$. The systematic uncertainties on the 39 
and 62.4 GeV data are similar in magnitude to those for 200 GeV 
presented in~\cite{Adare:2014fwh}. For integrated yield we treat 
every systematic uncertainty as \pt-correlated in the interest of 
consistency throughout the different data sets.

\begin{figure*}[bht]
\includegraphics[width=0.998\linewidth]{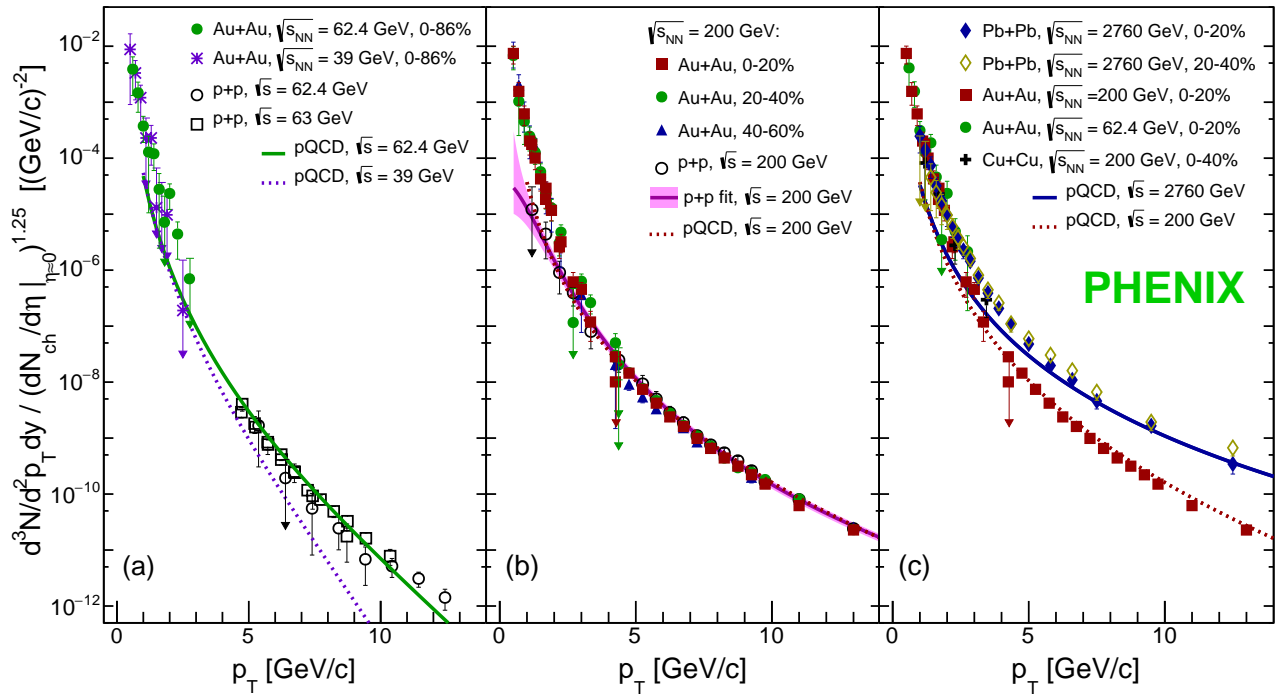}
\caption{
Direct photon spectra normalized by $(dN_{\rm ch}/d\eta)^{1.25}$ for 
\AB{Au}{Au} at 39 and 64.2 GeV (a) and (b) at 200 GeV 
\protect\cite{Adare:2014fwh}; panel (c) compares for different 
\AB{A}{A} systems at different \sqsn 
\protect\cite{Adam:2015lda,Adare:2018jsz}. Panels (a) and (b) also 
show \pp data 
\protect\cite{Adare:2014fwh,Angelis:1980yc,Angelis:1989zv,Akesson:1989hp}. 
All panels show pQCD calculations for the corresponding \sqs 
\protect\cite{Paquet:2015lta,Paquet:2017}. The errors shown are the 
quadratic sum of systematic and statistical uncertainties. 
Uncertainties on the $dN_{\rm ch}/d\eta$ are not included. }
\label{fig:yield}
\end{figure*} 

Figure~\ref{fig:yield} shows the invariant yield of direct photons 
normalized to $(dN_{\rm ch}/d\eta)^{1.25}$, this normalization is 
discussed below. Panel (a) shows \AB{Au}{Au} MB data at \sqsn = 
62.4 and 39 GeV, panel (b) gives \AB{Au}{Au} data in three 
centrality classes at 200 GeV, and panel (c) compares data from 
different beam energies and systems. Below 3 GeV/$c$ the 62.4 and 
39 GeV data show substantial direct photon yields, which are 
comparable in magnitude and spectral shape, albeit within large 
uncertainties. For 62.4 GeV we can also extract a direct photon 
signal for 0\%--20\% and 20\%--40\% centrality selection and find 
that the direct photon yield increases with centrality. All 
observations are similar to those already published for \AB{Au}{Au} 
collisions at \snn{200}~\cite{Adare:2014fwh}.

To compare data from different beam energies, collisions species, 
and collision centralities we use the measured charged particle 
multiplicity \dNch as measure of the system size at hadronization. 
For a fixed beam energy \dNch is roughly proportional \Npart. 
However, unlike \Npart, \dNch does not saturate but increases 
monotonically with beam energy for collisions of the same nuclei at 
the same impact parameter.

\begin{figure}[htb]
\includegraphics[width=1.0\linewidth]{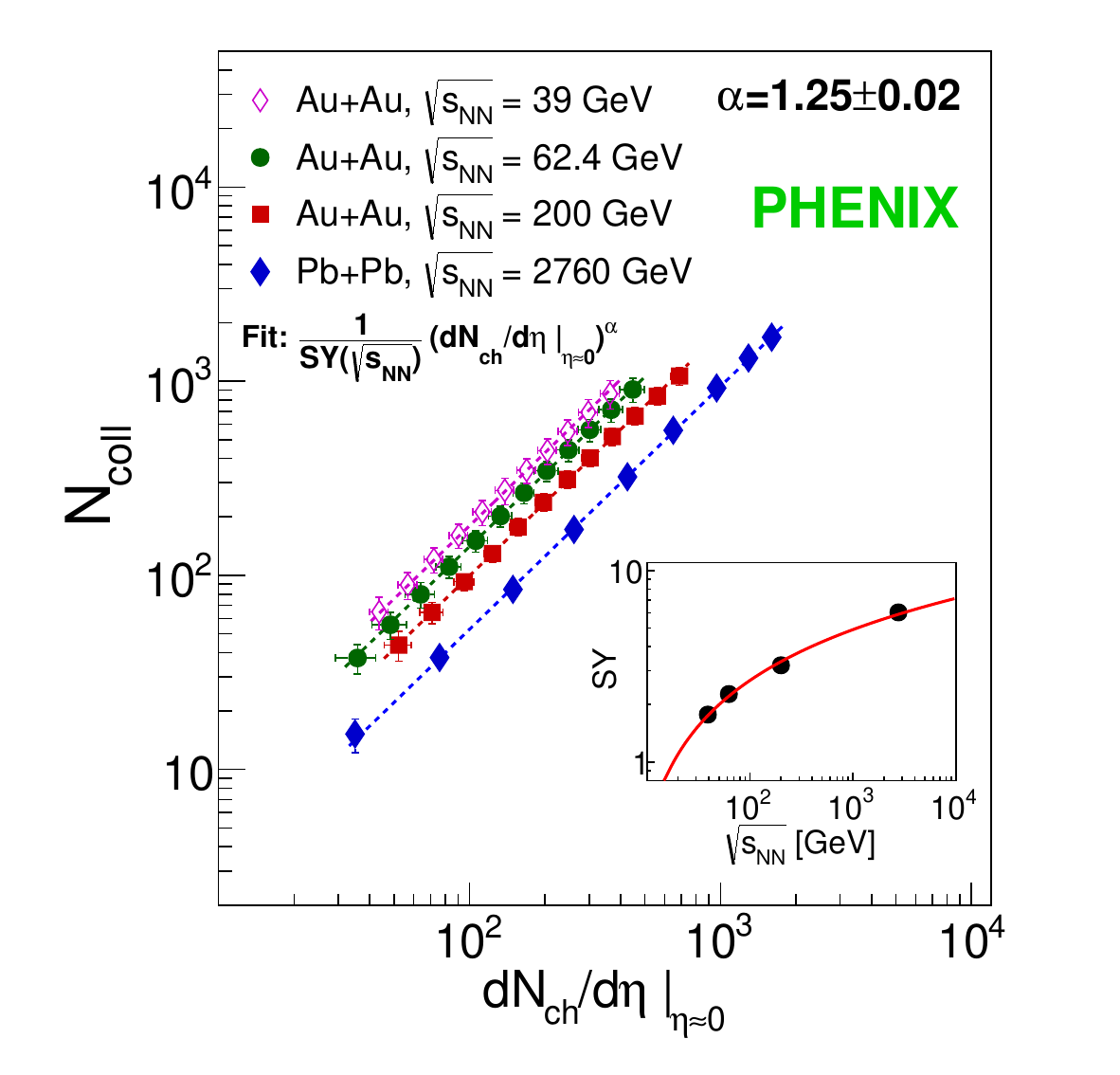}
\caption{
Number of binary collisions, \Ncoll , vs. \dNch, for four beam 
energies. The errors shown reflect the uncertainty of 
$N_{\rm coll}$ from the Glauber calculation. Fitting 
Eq.~(\ref{eqn:ncoll}) simultaneously to all data with a common 
$\alpha$ results in $\alpha$ = 1.25 and a \sqsn dependence $SY$ 
as shown in the text below Eq.~(\ref{eqn:ncoll}).}
\label{fig:Nch}
\end{figure}

Direct photon production at high \pt results from hard scattering, 
which at a fixed \sqsn scales with the number of binary collisions 
\Ncoll. We find that \Ncoll exhibits a remarkably simple relation 
with the \dNch that takes the form:
\begin{equation}
\Ncoll = \frac{1}{SY(\sqsn)} \times \Big( \frac{d\Nch}{d\eta}\Big)^\alpha
\label{eqn:ncoll}
\end{equation}
This is shown in Fig.~\ref{fig:Nch} where \Ncoll is plotted versus 
\dNch for different \sqsn.  PHENIX data are taken from 
\cite{Adare:2015bua} and ALICE data at \snn{2760} are from 
\cite{Aamodt:2010cz}. The exponent $\alpha$ is determined through a 
simultaneous fit to all data shown in Fig.~\ref{fig:Nch} and found 
to be $\alpha=1.25\pm0.02$. The specific yield $SY$ increases 
logarithmically with \sqsn as $SY(\sqsn)=(0.976 \pm 
0.054)\log{(\sqsn)}-(1.827{\pm}0.253)$.

Figure~\ref{fig:yield} depicts the direct photon yield for 
different beam energies and centralities normalized by 
$(dN_{\rm ch}/d\eta)^{1.25}$. In panel (b) three different centrality 
selections of \AB{Au}{Au} collisions at \snn{200} are shown 
together with data from \pp at the same beam energy. The normalized 
spectra from \AB{Au}{Au} are very similar for all three centrality 
selections. Above 3--4\,\gevc the normalized yield is the same as 
for \pp collisions and can be reproduced by perturbative quantum 
chromodynamics (pQCD) calculations with a renormalization and 
factorization scale of 
$\mu=0.5\pt$~\cite{Aurenche:2006vj,Paquet:2017}. Here the pQCD 
calculation was normalized to the experimental \dNch for \s{200} 
from~\cite{Patrignani:2016xqp}. Also shown on (b) is an empirical 
fit to the \pp data~\footnote{The fit function to \AB{p}{p} data 
originally used in \protect\cite{Adare:2008ab,Adare:2014fwh} was 
updated in~\protect\cite{Adare:2018jsz}. The parameters are $a = 
6.74\cdot 10^{-3}$, $b = 2.1$, $c=-3.3$. Systematic uncertainties 
also include possible shape variations at low \pt.} of the form 
$a(1+p_T^2/b)^c$~\cite{Adare:2018jsz}. Below 2--3\,\gevc the 
normalized yield in \AB{Au}{Au} collisions is significantly 
enhanced compared to that in \pp collisions, but follows the same 
scaling behavior with $(dN_{\rm ch}/d\eta)^{1.25}$ independent of 
centrality.

\begin{figure}[hbt]
\begin{minipage}{1.0\linewidth}
\includegraphics[width=0.998\linewidth]{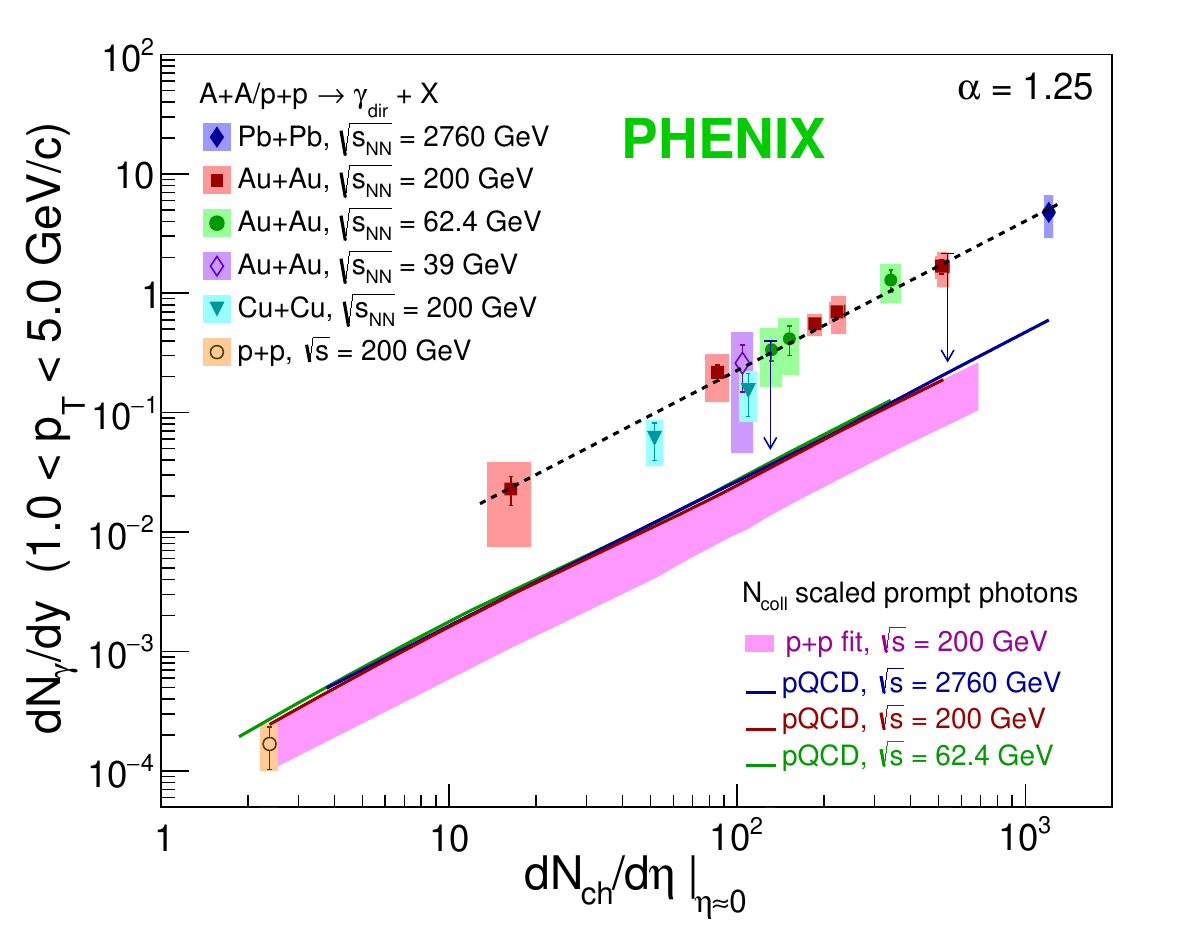}
\caption{Integrated direct photon yield ($p_T > 1.0$ \gevc) vs.
\dNch, for data sets shown in Fig.~\protect\ref{fig:yield}. The 
dashed line is a power law fit with a fixed slope of $\alpha=1.25$. 
The two upper limits correspond to the data in 20\%--40\% and 
40\%--80\% \AB{Pb}{Pb} collisions at \snn{2760}. The 
integrated yields of the fit to \pp data and of the pQCD 
calculations are shown as well, both scaled by 
\Ncoll~\protect\cite{Paquet:2015lta,Paquet:2017}.}
\label{fig:intyield1}
\end{minipage}
\begin{minipage}{1.0\linewidth}
\includegraphics[width=0.998\linewidth]{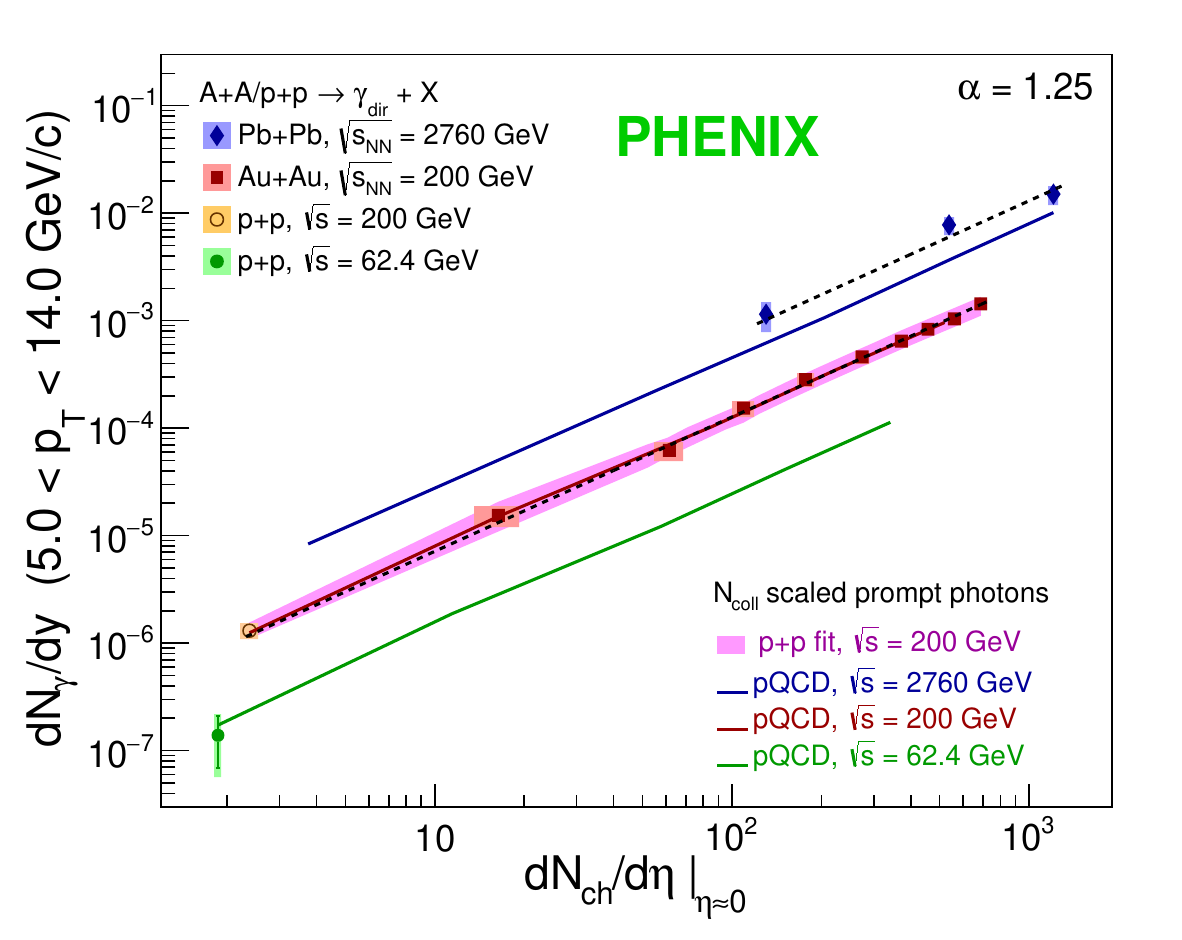}
\caption{Integrated direct photon yield ($p_T > 5.0$ GeV/$c$) vs. 
\dNch, for different data sets. The dashed lines show power law fits 
to the data with fixed slope of $\alpha=1.25$. Integrated yields
from pQCD calculations scaled by \Ncoll are also shown.}
\label{fig:intyield2}
\end{minipage}
\end{figure}

Panels (a) and (c) of Fig.~\ref{fig:yield} show that for \pt below 
2--3\,\gevc the same scaling with $(dN_{\rm ch}/d\eta)^{1.25}$ occurs 
for different \sqsn and collisions systems. Below 2 \gevc the 
spectra have very similar shape. We note that the apparent 
difference of the inverse slopes reported by PHENIX 
\cite{Adare:2014fwh} and ALICE \cite{Adam:2015lda} is largely due 
to the different fit ranges used~\footnote{When fitting the 
0\%--20\% \AB{Au}{Au} data at \snn{200} over the range 1.0 to 2.0 
\gevc, which overlaps the range 0.9 to 2.1 \gevc used by ALICE, 
instead of the original range of 0.6 to 2 \gevc deployed by PHENIX, 
we obtain an inverse slope of $279\pm32\pm10$ \mevc. This value is 
consistent with the value $297\pm12\pm41$ \mevc published by ALICE 
for the same centrality class for \AB{Pb}{Pb} at \snn{2760}.}.

At higher \pt the expected difference with \sqsn is observed. Like 
for \snn{200}, at high \pt the 2760\,GeV data are well reproduced 
by the pQCD calculation, though only above 5--6\,\gevc rather than 
3--4\,\gevc. Note that the extrapolated pQCD calculations for \pp 
at different \sqs seem to converge to the same normalized yield at 
low \pt, but at a tenth of the \AB{A}{A} yield.

We quantify direct photon emission by integrating the invariant 
yield above \pt=1.0\,\gevc and \pt=5.0\,\gevc. The integrals with 
the lower threshold will be dominated by excess low \pt photons 
unique to \AB{A}{A} collisions, while the integrals with the higher 
threshold are more sensitive to photons from initial hard 
scattering processes. The results are shown in 
Figs.~\ref{fig:intyield1}~and~\ref{fig:intyield2} as a function of 
\dNch. Also plotted are power-law functions 
$A(dN_{\rm ch}/d\eta)^{\alpha}$ with fixed $\alpha = 1.25$ and a 
normalization fitted to the data.

For \AB{A}{A} collisions the integrated yields for the 1.0 
\gevc threshold, shown in Fig.~\ref{fig:intyield1}, scale as ($7.140 \pm 
0.265) \times 10^{-4} \times (dN_{\rm ch}/d\eta)^{1.250}$. We find the 
same scaling if $\alpha$ is not constrained: $(8.300 \pm 1.680) 
\times 10^{-4} \times (dN_{\rm ch}/d\eta)^{1.225 \pm 0.034}$. The 
\AB{A}{A} points are compared to the integrated yield for \s{200} 
\pp obtained from the fit to the data, which is scaled with \Ncoll 
to the corresponding \dNch for each \snn{200} \AB{A}{A} point. The 
width of the band is given by the combined uncertainties on the fit 
function and \Ncoll. It is parallel to the \AB{A}{A} trend but 
lower by about an order of magnitude. Also shown are the scaled 
integrated yields from pQCD calculations for \s{62.4, 200, and 
2760}, consistent with the band independent of beam energy.

For the \pt threshold of 5\,GeV$/c$ the integrated yields from 
\AB{Au}{Au} and \pp at 200 GeV follow the same 
$(dN_{\rm ch}/d\eta)^{1.25}$ trend, and are described by the pQCD 
calculation. The 2760\,GeV data are also consistent with 
$(dN_{\rm ch}/d\eta)^{1.25}$ but show a significantly higher yield than 
at 200 GeV data at the same \dNch. The \Ncoll scaled pQCD 
calculation is about 30\% below the data, which may not be 
significant considering the 25\% systematic uncertainty on the 
calculation.

While the functional form $A(\dNch)^{\alpha}$ describes the 
integrated direct photon yields well, it is not unique. For 
instance the data can be equally well fitted by $A(\dNch) + 
B(\dNch)^{4/3}$ \cite{Feinberg:1976ua}. For the data in 
Fig.~\ref{fig:intyield1} this fit results in parameters $A = (8.68 
\pm 3.06) \cdot 10^{-4}$ and $B = (3.09 \pm 0.45) \cdot 10^{-4}$. 
The important point is that \AB{A}{A} data from different 
centralities and a wide range of collision energies can be 
empirically described in terms of \dNch with just two parameters, 
suggesting some fundamental commonality in the underlying physics.

There are two main conclusions from the analyses presented in this 
paper. (i) At a given beam energy the direct photon yield scales 
with $\dNch^{1.25}$ or \Ncoll for all observed \pt. There seems to 
be no qualitative change in the photon sources and/or their 
relative contributions for different collision centrality or system 
size. (ii) From \snn{39 to 2760} the same scaling is observed for 
\pt$<$ 2 GeV/$c$. This suggests that the main sources contributing 
to this \pt range are very similar also across beam energies.

If thermal radiation is the source of low \pt direct photons, the 
similarity at the same \dNch across beam energies and centralities 
for \pt$\lesssim$ 2 GeV/$c$, suggests that the bulk of the matter 
that emits the radiation is similar in terms of temperature and 
space time evolution. This would be natural, if most of the photons 
are emitted near the transition from QGP to hadrons.

While at high \pt the scaled yields in \pp and \AB{A}{A} are 
identical, at low \pt they differ by a factor of 10. This implies 
that there must be a transition from the small \pp yield to the 
enhanced \AB{A}{A}-like low \pt yields in the \dNch range of 
${\approx}2$ to 20, which will be accessible with the data taken by 
PHENIX with small systems $p$$+$Au, $d$$+$Au, and $^3$He$+$Au.



\begin{acknowledgments}

We thank the staff of the Collider-Accelerator and Physics
Departments at Brookhaven National Laboratory and the staff of
the other PHENIX participating institutions for their vital
contributions.  We acknowledge support from the 
Office of Nuclear Physics in the
Office of Science of the Department of Energy,
the National Science Foundation, 
Abilene Christian University Research Council, 
Research Foundation of SUNY, and
Dean of the College of Arts and Sciences, Vanderbilt University 
(U.S.A),
Ministry of Education, Culture, Sports, Science, and Technology
and the Japan Society for the Promotion of Science (Japan),
Conselho Nacional de Desenvolvimento Cient\'{\i}fico e
Tecnol{\'o}gico and Funda\c c{\~a}o de Amparo {\`a} Pesquisa do
Estado de S{\~a}o Paulo (Brazil),
Natural Science Foundation of China (People's Republic of China),
Croatian Science Foundation and
Ministry of Science and Education (Croatia),
Ministry of Education, Youth and Sports (Czech Republic),
Centre National de la Recherche Scientifique, Commissariat
{\`a} l'{\'E}nergie Atomique, and Institut National de Physique
Nucl{\'e}aire et de Physique des Particules (France),
Bundesministerium f\"ur Bildung und Forschung, Deutscher Akademischer 
Austausch Dienst, and Alexander von Humboldt Stiftung (Germany),
J. Bolyai Research Scholarship, EFOP, the New National Excellence
Program ({\'U}NKP), NKFIH, and OTKA (Hungary),
Department of Atomic Energy and Department of Science and Technology 
(India),
Israel Science Foundation (Israel), 
Basic Science Research Program through NRF of the Ministry of 
Education (Korea),
Physics Department, Lahore University of Management Sciences (Pakistan),
Ministry of Education and Science, Russian Academy of Sciences,
Federal Agency of Atomic Energy (Russia),
VR and Wallenberg Foundation (Sweden), 
the U.S. Civilian Research and Development Foundation for the
Independent States of the Former Soviet Union, 
the Hungarian American Enterprise Scholarship Fund,
the US-Hungarian Fulbright Foundation,
and the US-Israel Binational Science Foundation.

\end{acknowledgments}



%
 
\end{document}